\documentclass[11pt]{article}
\usepackage[leqno]{amsmath}
\usepackage{amssymb, amsthm, amsfonts, graphicx, enumerate, mathrsfs, bbm, amsmath, stackrel}

\newtheoremstyle{nonum}{}{}{\itshape}{}{\bfseries}{.}{ }{\thmnote{#3}}

\newtheorem{thm}{Theorem}
\newtheorem{clm}[thm]{Claim}
\newtheorem{lem}[thm]{Lemma}
\newtheorem{prop}[thm]{Proposition}

\theoremstyle{definition}
\newtheorem*{prf}{Proof}
\newtheorem*{rem}{Remark}

\newcommand{\R}{\mathbb R}
\newcommand{\Z}{\mathcal Z} 
\newcommand{\A}{\mathcal A} 
\newcommand{\C}{\mathcal C} 

\begin{document}

\title{On the zone of the boundary of a convex body\thanks{Work on this paper has been supported by Grant 338/09 from the Israel Science Fund and by the Hermann Minkowski MINERVA Center for Geometry at Tel Aviv University.}}
\author{Orit E. Raz\thanks{School of Computer Science, Tel Aviv University, Tel Aviv 69978, Israel; \texttt{oritraz@post.tau.ac.il}}}
\maketitle

\begin{abstract}
We consider an arrangement $\A$ of $n$ hyperplanes in $\R^d$ and the zone $\Z$ in $\A$ of the boundary of an arbitrary convex set in $\R^d$ in such an arrangement. We show that, whereas the combinatorial complexity of $\Z$ is known only to be $O\left(n^{d-1}\log n\right)$ \cite{APS}, the outer part of the zone has complexity $O\left(n^{d-1}\right)$ (without the logarithmic factor). Whether this bound also holds for the complexity of the inner part of the zone is still an open question (even for $d=2$).
\end{abstract}

\section{Introduction}

A collection $H$ of $n$ hypersurfaces in $\R^d$ decomposes $\R^d$ into open cells of dimension $d$ and into (relatively open) faces whose dimension $k$ varies between 0 and $d-1$. Each of these faces and cells is a maximal connected relatively open set contained in a fixed subset of $H$ and avoiding all other hypersurfaces. We refer to $(d-1)$-dimensional faces as \emph{facets}, to 0-dimensional faces as \emph{vertices}, to 1-dimensional faces as \emph{edges}, and to faces of dimension $k$, where $2\leq k\leq d-2$, simply as \emph{$k$-faces}. These cells and faces define a cell complex, called the \emph{arrangement} $\A=\A(H)$ of $H$ (see \cite{OT}). 

The \emph{complexity} of a cell in $\A$ is defined to be the number of faces (of all dimensions) that are contained in the closure of the cell. For a set $\sigma\subset\R^d$, the \emph{zone} of $\sigma$ is defined to be the set of all cells in $\A$ that $\sigma$ intersects. The \textit{complexity} of a zone is the sum of the complexities of the cells in the zone. For general hypersurfaces, studying the complexity of a zone in an arrangement is closely related to the problem of analyzing the maximum possible complexity of a single cell (see \cite[Section 5.3]{SA}).

In this paper we focus on the special case where $H$ is a collection of $n$ hyperplanes in $\R^d$.

\paragraph{Planar arrangements.}

It is a trivial fact that the number of edges or vertices of a single cell in an arrangement of $n$ lines in the plane is at most $n$. A linear bound on the complexity of a cell in an arrangement of halflines is also known (see Alevizos et al.~\cite{ABP}). On the other hand, a result of Wiernik and Sharir \cite{WS} on the lower envelopes of segments implies that a single cell in an arrangement of $n$ line segments in the plane can have $\Omega\left(n\alpha(n)\right)$ vertices in the worst case, where $\alpha(n)$ is the inverse Ackermann function. A matching upper bound has been proved by Pollack et al.~\cite{PSS}.

\begin{figure}
\centering
\includegraphics[width=0.8\textwidth]{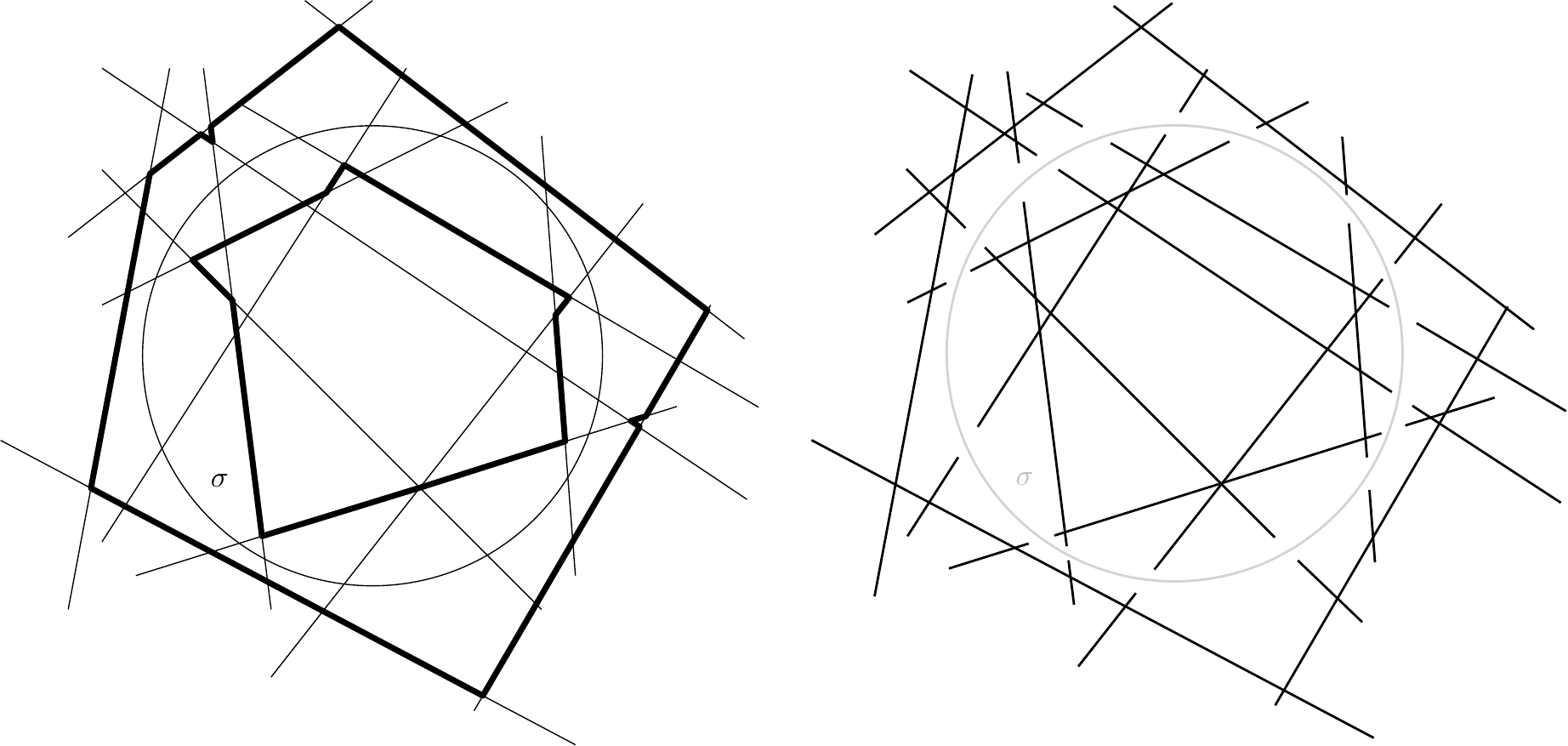}
\caption{Left: The zone of a circle $\sigma$ in a line arrangement (consisting all the cells between the two highlighted cycles). Right: Turning the zone into a single cell.}
\label{fig:zonecell}
\end{figure}

To demonstrate the relation of these results to the zone problem, consider the following problem: Let $L$ be a collection of $n$ lines in the plane and let $\sigma$ be a given circle drawn in the plane. What is the complexity of the zone of $\sigma$ in $\A(L)$? To solve this problem we use the following reduction (following \cite{SA}): For each intersection point of $\sigma$ with a line $l\in L$, we split $l$ into two pieces and leave an arbitrarily small gap between these pieces. In this manner, all faces of the zone of $\sigma$ are merged into one cell, now of an arrangement of (at most $3n$) lines, halflines and line segments (see Figure \ref{fig:zonecell}). The aforementioned results imply that the complexity of the zone in question is $O\left(n\alpha(n)\right)$. More precisely, the part of the zone outside the circle $\sigma$ has complexity $O\left(n\right)$ (because, if we erase the portions of the lines inside the disk $D$ bounded by $\sigma$, the outer zone becomes a portion of a single cell in an arrangement of $O\left(n\right)$ halflines and lines). On the other hand the complexity of the inner part of the zone (which becomes a portion of a single cell in an arrangement of $O\left(n\right)$ line segments if we erase the portions of the lines outside $D$) is known only to be $O\left(n\alpha(n)\right)$. Whether this complexity bound (of the inner part) is tight, is a major open problem.

A major role in the derivation of these results is played by \emph{Davenport-Schinzel sequences}, combinatorial structures that arise in the analysis of lower envelopes of collections of univariate functions. Analyzing the envelope of a collection of multivariate functions appears to be considerably harder than the univariate case and the known bounds on its complexity are not as sharp as in the planar case (for more details see the book \cite{SA} and the survey paper \cite{AS}). 

\paragraph{Higher dimensions.}

A fundamental result on \emph{hyperplane} arrangements is the (standard) Zone Theorem \cite{ESS, Ede}, in which $\sigma$ is assumed to be a hyperplane distinct from those in $H$. It asserts that the zone of a hyperplane in an arrangement of $n$ hyperplanes in $\R^d$ has complexity $\Theta\left(n^{d-1}\right)$, where the constant of proportionality depends on $d$. In \cite{APS}, the Zone Theorem is extended to cases where $\sigma$ is either an algebraic variety or the boundary of an arbitrary convex set. It is shown in \cite{APS} that the complexity of the zone in these settings is $O\left(n^{d-1}\log n \right)$, where the constant of proportionality depends on $d$ and, in the former case, also on the degree of $\sigma$.

In this paper, we consider hyperplane arrangements in $\R^d$ and the zone of the boundary of an arbitrary convex set in $\R^d$. We show (in Theorem \ref{main}) that, similar to the planar case, taking into account only the outer part of the zone (precise definition given below) has complexity $O\left(n^{d-1}\right)$ (without the logarithmic factor). As in the planar case, whether this bound also holds (in the worst case) for the complexity of the inner part of the zone is still an open question.

\vspace{8mm}
\noindent\emph{Acknowledgment.} I would like to thank my advisor Prof. Micha Sharir for his help, advice, useful discussions and support.

\section{The outer complexity of the zone}

Let $H$ be a collection of $n$ hyperplanes in $\R^d$ and let $K$ be a closed convex set in $\R^d$. We denote the zone of $\partial K$ in $\A$ as $\Z=\Z(K,H)$, and define the \emph{outer complexity} of $\Z$ (with respect to $K$) to be 
\begin{equation*}
\C(\Z)=\sum_{C\in\Z}\sum_{i=0}^{d-1}f_{K^c}^i(C),
\end{equation*}
where $K^c:=\R^d\setminus K$ and $f_{K^c}^i(C)$ is the number of $i$-faces on the boundary of the cell $C$ that are \emph{contained} in $K^c$.

We assume that $H$ and $K$ are in \emph{general position}, i.e., any $j\leq d$ hyperplanes in $H$ intersect in a $(d-j)$-flat, no $d+1$ hyperplanes have a point in common, and $\partial K$ is not tangent to any flat formed by the intersection of any number $j\leq d$ hyperplanes in $H$ (in particular, $\partial K$ does not meet any vertex of $\A$). This assumption involves no loss of generality. This can be proved using a standard perturbation argument: Displacing the hyperplanes of $H$ slightly will put $K$ and $H$ in general position, and can only increase the complexities of the outer part of the cells in $\Z$ (see \cite[pp. 78--83]{Gru}).

Note that a $k$-face $f$ in $\A$ lies in $d-k$ hyperplanes of $H$ and is on the boundary of $2^{d-k}$ cells of $\A$. More than one of these cells can belong to $\Z$, and thus the contribution of $f$ to $\C(\Z)$ can be larger than one. It is therefore more convenient to consider pairs $(f,C)$, where $C\in\Z$ and $f$ is a $k$-face in $\A$ that lies on the boundary of $C$. Each such pair contributes at most one to the count (it contributes when $f\subset K^c$). Following a similar notation in \cite{APS}, we call such a pair a \textit{$k$-border}.

The main result of this paper is the following theorem.
\begin{thm}\label{main}
Let $H$ be a collection of $n$ hyperplanes in $\R^d$, let $K\subset\R^d$ be a closed convex set, and put $\Z=\Z(K,H)$. Then
\begin{equation}
\C(\Z)\leq Adn^{d-1},
\end{equation}
where $A$ is an absolute constant independent of $d$.
\end{thm}

\begin{rem}
In the definition of the outer complexity of $\Z$ we ignore, not only faces that lie fully inside of $K$, but also faces that intersect $\partial K$. It is natural though, to include faces of the latter kind in the complexity of the outer zone, and it is not hard to handle such faces, and to show that their total number is $O\left(n^{d-1}\right)$. Indeed, as is shown in \cite[Lemma 2.3]{APS}, the number of ($d$-dimensional) cells in $\Z$ is $O\left(n^{d-1}\right)$. To bound the number of $k$-faces that intersect $\partial K$, one can simply restrict $\A$ to each of the $k$-flats, formed by the intersection of some $d-k$ hyperplanes of $H$, and apply this bound on the number of cells to the restricted (lower-dimensional) arrangements. A more detailed discussion of this issue is given at the end of the analysis.
\end{rem}

\noindent{\bf{Proof of Theorem \ref{main}.}}
For each $0\leq i< d$, we put 
\begin{equation*}
\tau_i(K,H):= \sum_{C\in\Z(K,H)}f^i_{K^c}(C);
\end{equation*}
that is, $\tau_i(K,H)$ is the number of $i$-borders $(f,C)$ such that $C$ is in $\Z$ and $f$ is (an $i$-face on the boundary of $C$ which is) contained in $K^c$.
By definition,
\begin{equation}
\C(\Z)=\sum_{i=0}^{d-1}\tau_i(K,H),
\end{equation}
and so it is sufficient to bound each of the quantities $\tau_i(K,H)$.

For $d=1$, $K$ is necessarily an interval and $H$ is a set of points, and we trivially have $\tau_0(K,H)\leq 2$.

\begin{figure}
\centering
\includegraphics[scale=0.5]{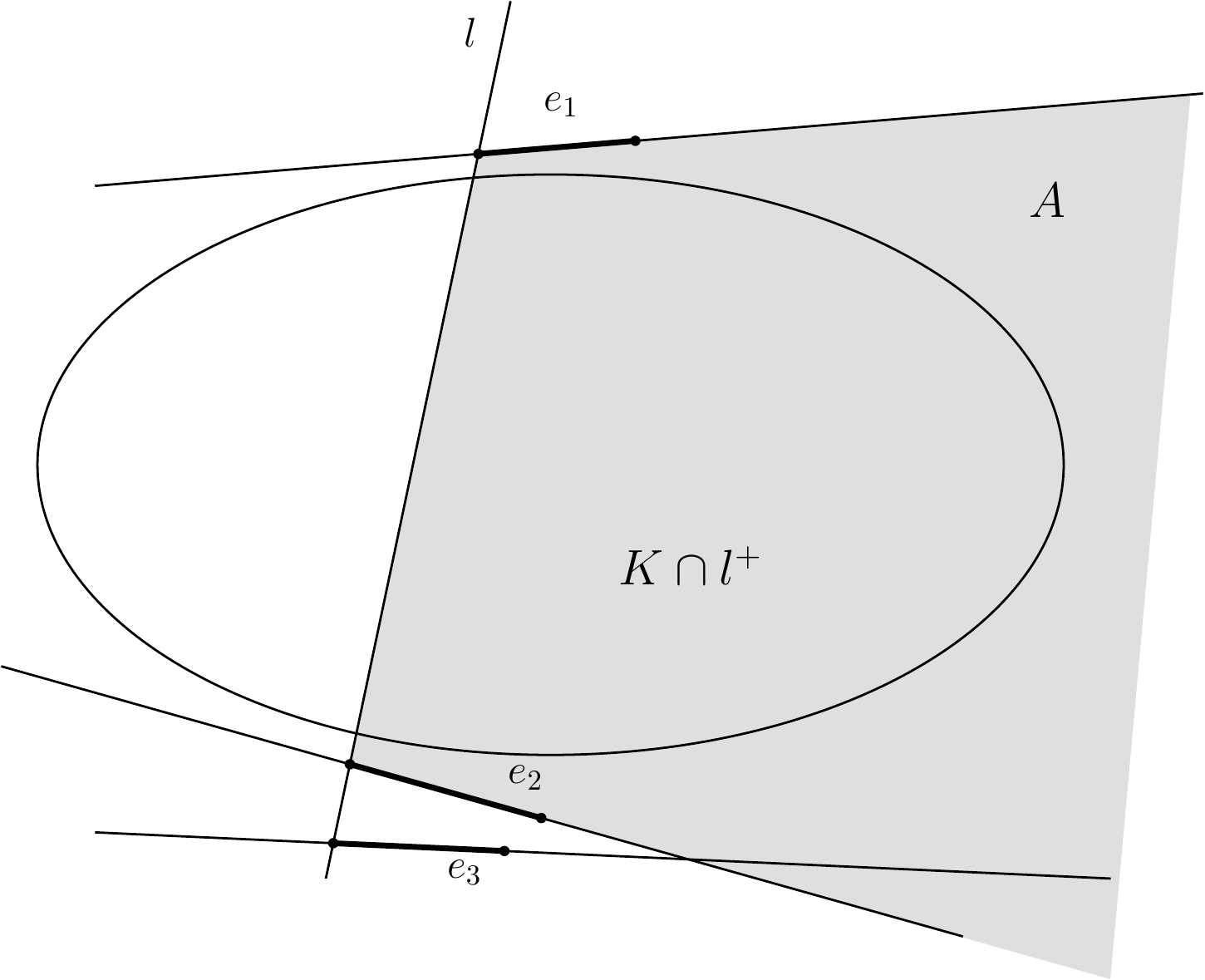}
\caption{Illustration for the proof of Lemma \ref{lem:dim2}.}
\label{fig:lem22}
\end{figure}

The case $d=2$ is treated in the following lemma.
\begin{lem}\label{lem:dim2}
For $d=2$ we have
\begin{equation}
\tau_0(K,H)\leq 12n\;\text{ and }\; \tau_1(K,H)\leq 4n.
\end{equation}
\end{lem}
\begin{prf}
To bound $\tau_1(K,H)$, consider an edge $e$ contained in $K^c$ and incident to a cell in $\Z$, and let $h\in H$ be the line that supports $e$. We use the property that no such edge $e$ is incident to more than one cell in $\Z$ (this is a special case of a more general property established in Claim \ref{clm:nopop} below), and thus the number of these edges is the same as the number of 1-borders we are interested in. We may assume that $e$ has at least one vertex, since the lines in $H$ are in general position, and we may assume that $n\geq 2$. 

Suppose first that $e$ has two vertices $u_1, u_2$. For $j=1,2$, we let $l_j\in H$ be the second line (besides $h$) defining the vertex $u_j$ and let $l_j^+$ denote the halfplane bounded by $l_j$ that contains $e$. Then there exists $j_0\in\{1,2\}$ such that the halfline $h\cap l_{j	_0}^+$ does not intersect $\partial K$ (and $K$); indeed, if this were not the case then by the convexity of $K$ the edge $e$ would have intersected $K$ as well (in fact would be contained in $K$), contrary to our assumption that $e\subset K^c$. We charge $e$ to the pair $(l_{j_0}, h\cap l_{j_0}^+)$. If $e$ has only one vertex $u$, we let $l$ be the second line (besides $h$) defining $u$ and charge $e$ to $(l, e)$; here too, $e$ is a halfline which (trivially) does not intersect $\partial K$.

We claim that for every line $l\in H$, there are at most four pairs of the form $(l,r)$ charged by edges $e$ as described above (that is, $l$ is a line of $H$ and $r$ is a halfline of another line that ends on $l$). More precisely, we will show that there are at most two pairs $(l,r)$ of this kind such that $r\subseteq l^+$, for each of the two halfplanes $l^+$ bounded by $l$. Indeed, suppose that there were three such pairs $(l,r_j)$, charged by three different edges as before and with $r_j\subset l^+$, for $j=1,2,3$. Let $h_j\in H$, for $j=1,2,3$, be the line that supports $r_j$ and let $h_j^+$ be the halfplane bounded by $h_j$ that contains $K\cap l^+$. Since the halflines lie on the same side of $l$, their containing lines $h_j$ are all distinct. Moreover, since the $r_j$'s are disjoint from $K$, the halfplanes $h_j^+$ are well defined. Consider the intersection $A:=l^+\cap h_1^+ \cap h_2^+ \cap h_3^+$ (see Figure \ref{fig:lem22} for an illustration). Since $A$ is convex, $l$ supports at most one edge on the boundary of $A$ and thus for some $j_1\in\{1,2,3\}$, the vertex $v:=l\cap r_{j_1}$ is not of the boundary of $A$. Let $e$ be the edge supported by $r_{j_1}$ incident to $v$. By assumption $e$ is on the boundary of a cell $C\in \Z$. But then, $C\subset l^+\cap A^c$, and hence it cannot intersect $\partial K$, which is a contradiction. Thus,
\begin{equation}\label{2}
\tau_1(K,H)\leq 4n.
\end{equation}

Finally, observe that
\begin{equation}
\tau_0(K,H)\leq 2\tau_1(K,H)+ 4n,
\end{equation}
since for every 0-border $(v,C)$ with $v\in K^c$, $v$ is incident to an edge $e$ on the boundary of $C$ such that either $e$ is fully contained in $K^c$ or $e$ intersects $\partial K$. Each line generates at most two segments of the latter kind, which can generate at most four $0$-borders. Together with inequality (\ref{2}), this gives
\begin{equation}
\tau_0(K,H)\leq 12n,
\end{equation}
as asserted.
$\square$
\end{prf}

For $d\geq 3$ we use an inductive argument, similar to the one employed in \cite{APS, ESS}.

\begin{prop}\label{prop:tauid}
For $d\geq 3$ and $2\leq i< d$, we have
\begin{equation}\label{1}
\tau_i(K,H)\leq 4(d-i)!\binom{n}{d-i}n^{i-1}.
\end{equation}
\end{prop}
\noindent (Note that the bound on the right-hand side is at most $4n^{d-1}$.)
\begin{prf}
We use induction on the dimension $d$, where the induction base, for $d=2$, has already been established in Lemma \ref{lem:dim2}. For the induction step, fix $d\geq 3$ and assume that (\ref{1}) holds for all $d'< d$ and $2\leq i< d'$.

\begin{lem}\label{lem:rec}
For each $1\leq i< d$, we have
\begin{equation}
(n-d+i)\tau_i(K,H)\leq \sum_{h\in H} \tau_i(K,H\setminus\{h\}) + \sum_{h\in H} \tau_{i-1}(K\cap h, H\cap h),
\end{equation}
where $H\cap h:=\{h'\cap h\;\;|\; h'\in H\setminus\{h\}\}$, viewed as a collection of $n-1$ hyperplanes (of dimension $d-2$) in $h$ (which is a translated and rotated copy of $\R^{d-1}$).
\end{lem}

We defer the proof of Lemma \ref{lem:rec} for later and assume for the moment that the lemma holds. Together with the induction hypothesis (applied with the parameters $d-1, i-1$ and $n-1$), the lemma implies
\begin{align*}
(n-d+i)\tau_i(K, H)
& \leq \sum_{h\in H} \tau_i(K, H\setminus\{h\}) + \sum_{h\in H} \tau_{i-1}(K\cap h, H\cap h)\\
& \leq \sum_{h\in H} \tau_i(K, H\setminus\{h\}) + n\cdot 4(d-i)!\binom{n-1}{d-i}(n-1)^{i-2}\\
&= \sum_{h\in H} \tau_i(K, H\setminus\{h\}) + 4(d-i)!(n-d+i)\binom{n}{d-i}(n-1)^{i-2}.
\end{align*}
For $i$ fixed, we use induction on $n$. For $n=|H|=d-i$ there is exactly one $i$-face $f$ in $\A(H)$ and at most $2^{d-i}$ $i$-borders $(f,C)$ with $C\in\Z$, thus (\ref{1}) holds. For larger values of $n$ we have, by the induction hypothesis,
\begin{align*}
\tau_i(K, H)
& \leq \frac{1}{n-d+i}\sum_{h\in H} \tau_i(K, H\setminus\{h\}) + 4(d-i)!\binom{n}{d-i}(n-1)^{i-2}\\
& \leq \frac{4n\cdot (d-i)!}{n-d+i}\binom{n-1}{d-i}(n-1)^{i-1} + 4(d-i)!\binom{n}{d-i}(n-1)^{i-2}\\
& = 4(d-i)!\binom{n}{d-i}n^{i-1}\left(\frac{n-1}{n}\right)^{i-2}\left[1-\frac{1}{n}+\frac{1}{n}\right]\\
& \leq 4(d-i)!\binom{n}{d-i}n^{i-1}\left(1-\frac{1}{n}\right)^{i-2}\\
& \leq 4(d-i)!\binom{n}{d-i}n^{i-1}\;\;\;\text{(because $i\geq 2$).}
\end{align*}
This establishes the induction step and thus completes the proof of Proposition \ref{prop:tauid}. $\square$
\end{prf}

\begin{figure}
\centering
\includegraphics[scale=0.8]{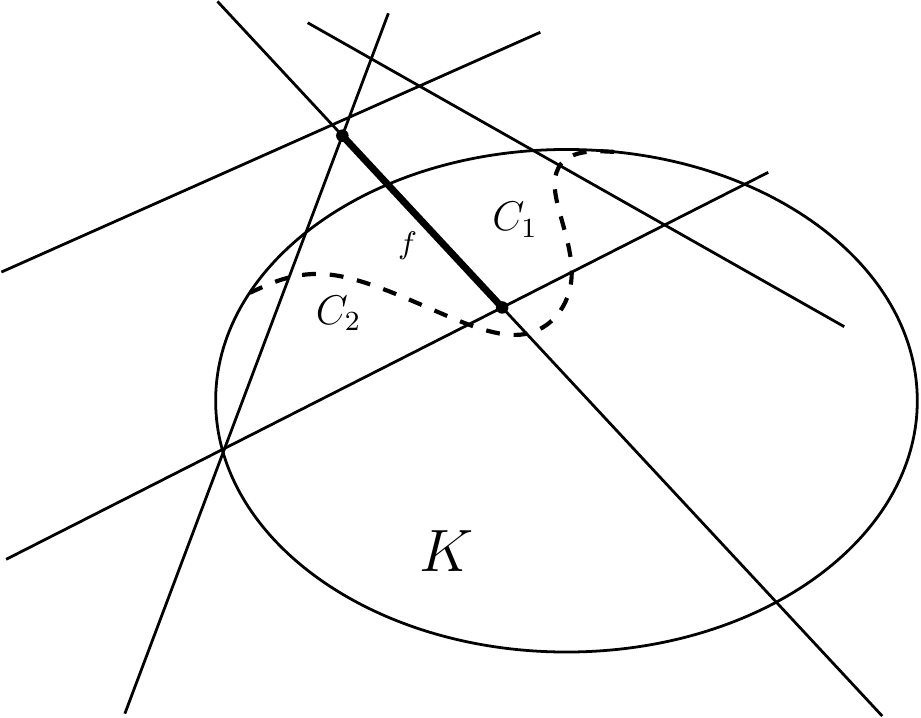}
\caption{Illustration for the proof of Claim \ref{clm:nopop}. For a convex set $K$ any facet $f$ that is incident to two cells $C_1,C_2\in \Z$ must intersect $\partial K$. This property does not have to hold when $K$ is non-convex, as is indicated by the dashed line.}
\label{fig:popfacet}
\end{figure}

We next provide the proof of Lemma \ref{lem:rec}. We need the following simple property (which is illustrated in Figure \ref{fig:popfacet}), whose planar case has already been used in the proof of Lemma \ref{lem:dim2}.
\begin{clm}\label{clm:nopop}
Let $f$ be a facet in $\A$ such that $f\cap K^c\neq\emptyset$. Then either $f\cap\partial K\neq\emptyset$ or $f$ is on the boundary of at most one cell in $\Z$.
\end{clm}
\begin{prf}
Let $f$ be a facet in $\A$ such that $f\cap K^c\neq\emptyset$ and assume that $f$ is on the boundary of two distinct cells $C_1, C_2$ in $\Z$. Observe first that $C_1\cup C_2\cup f$ is convex--- removal of the hyperplane supporting $f$ merges $C_1, C_2$, and (the missing) $f$ into a single, necessarily convex, cell.

Suppose for the sake of contradiction that $f$ does not intersect $\partial K$. Since $f\cap K^c\neq\emptyset$ this implies that $f\subset K^c$. Now let $x_1\in C_1\cap \partial K$ and $x_2\in C_2\cap\partial K$; $x_1,x_2$ exist since $C_1, C_2$ are both in $\Z$. Then the segment $x_1x_2$ connecting the points $x_1$ and $x_2$ intersects $h_f$, because $C_1$ and $C_2$ lie on different sides of it. But $C_1\cup C_2\cup f$ is convex and so $x_1x_2$ intersects $f$ (because $f=h_f\cap (C_1\cup C_2\cup f)$).

However, since $K$ is a convex (and closed) set, every segment connecting two points on its boundary must be contained in $K$. But we have shown that $f\subset K^c$, which is a contradiction. Thus $f$ intersects $\partial K$.~$\square$
\end{prf}

We note that the convexity of $K$ is crucial for Claim \ref{clm:nopop} to hold. An instance where Claim \ref{clm:nopop} fails for a non-convex $K$ is depicted in Figure \ref{fig:popfacet} (where the dashed line depicts the relevant portion of $\partial K$).

\noindent{\bf{Proof of Lemma \ref{lem:rec}.}} Remove an arbitrary hyperplane $h$ from $H$ and consider $\A':=\A(H\setminus\{h\})$. Our goal is to bound the increase in the number of $i$-borders in $\A(H)$ that \emph{are not contained in $h$}, over their number in $\A(H\setminus\{h\})$. For this, let $(e',C)$ be an $i$-border in $\A'$, such that $e'\subset\partial C$ is an $i$-face contained in $K^c$. Now reinsert $h$ into the arrangement. We are only interested in the situation where $(e',C)$ is split by $h$ into two $i$-borders in $\A:=\A(H)$ and both of the cells of these borders are in $\Z$, because only then will our count go up. Note that in this case $f:=C\cap h$ is a facet in $\A$ which lies on the boundary of the two zone cells $C_1:=C\cap h^+$ and $C_2:=C\cap h^-$, where $h^+, h^-$ denote the two open halfspaces bounded by $h$. By Claim \ref{clm:nopop}, $f$ intersects $\partial K$ (this follows since $e'\subset K^c$ so $e'\cap f\subset K^c$ and this implies that $f$ meets $K^c$). Thus $(e,f)$, where $e:=e'\cap h$, is an $(i-1)$-border in the $(d-1)$-dimensional arrangement $\A(H\cap h)$ (within $h$) with respect to $\Z(K\cap h, H\cap h)$. Moreover, $e$ is contained in $K^c$ (because $e'$ is). Thus we may charge the split of $(e',C)$ to $(e,f)$ in a unique fashion.

To sum up, if we consider only $i$-borders with $i$-faces not supported by $h$, the insertion of $h$ into $\A'$ increases the count by at most $\tau_{i-1}(K\cap h, H\cap h)$. Summing these bounds over all hyperplanes $h\in H$, and using the fact that the number of hyperplanes not supporting a specific $i$-face is $n-d+i$, complete the proof of the lemma. $\square$

To complete the argument we need to extend the bound given in Proposition \ref{prop:tauid} to the cases $i=0$ and $i=1$, which is taken care of in the following lemma.
\begin{lem}\label{lem:tau01}
For $d\geq 3$, we have
\begin{equation}
\tau_0(K,H)= O\left(n^{d-1}\right)\;\;\text{ and }\;\;\tau_1(K,H)= O\left(dn^{d-1}\right),
\end{equation}
where the constants of proportionality are independent of $d$.
\end{lem}
\begin{prf}
First, observe that
\begin{equation}\label{eq:tau1}
\tau_1(K,H)\leq \frac{d}{2}\tau_0(K,H).
\end{equation}
Indeed, let $(e,C)$ be a 1-border such that $C$ is a zone cell and $e$ is an edge on $\partial C$ contained in $K^c$. Obviously (using the general position assumption), the vertices $v_1,v_2$ incident to $e$ are also inside $K^c$, and thus the 0-borders $(v_1,C), (v_2,C)$ contribute both to the count in $\tau_0(K,H)$; we charge $(e,C)$ to them. Since every 0-border can be charged by at most $d$ 1-borders, the inequality (\ref{eq:tau1}) follows.

To bound $\tau_0(K,H)$, we will consider 3-faces $P$ in $\A$ (that bound cells of $\Z$) and use Euler's formula to bound the number of vertices incident to $P$ in terms of the number of 2-faces incident to $P$. This is a standard technique, already used in \cite{APS}, and it produces (as we show below) a bound on $\tau_0(K,H)$ using the bound that we already have on $\tau_2(K,H)$. However, since in our setting not all the 2-faces on the boundary of $P$ are necessarily contained in $K^c$, and therefore not all of them necessarily contribute to the count in $\tau_2(K,H)$, one needs to be somewhat more careful using Euler's formula. To address this technical issue, we separate our analysis into the following two cases.

Let $(v,C)$ be a 0-border such that $C$ is a zone cell and $v$ is a vertex on $\partial C$ contained in $K^c$. Consider first the the case where there exists a 2-border $(f,C)$, such that $v$ is incident to $f$, and the 2-face $f$ intersects $\partial K$. In this case, pick such a 2-border $(f,C)$ and charge $(v,C)$ to the pair $(v,f)$. Next, consider the unique plane $F$ that supports $f$, formed by the intersection of some $d-2$ hyperplanes in $H$. Then $(v,f)$ is a 0-border in the 2-dimensional arrangement $\A':=\A(H\cap F)$, where $H\cap F:=\{h\cap F\;|\; h\in H \;\text{ and }\; F\nsubseteq h\}$ is a collection of $n-d+2$ lines in $F$. Moreover, $f$ is a zone cell in $\Z(K\cap F,H\cap F)$ and $v$ is contained in $K^c\cap F=(K\cap F)^c\cap F$. By Lemma \ref{lem:dim2}, there are at most $12n$ 0-borders on $F$ of this form. Going over all the $\binom{n}{d-2}$ planes, each formed by the intersection of some $d-2$ hyperplanes in $H$, gives a bound of at most $12n\binom{n}{d-2}$ on the number of such pairs $(v,f)$.

Hence, the number of 0-borders $(v,C)$ in $\A$ such that $v$ is contained in $K^c$ and is incident to a 2-face on the boundary of $C$ that intersects $\partial K$, is at most
\begin{equation}\label{eq1}
12\cdot 2^{d-2}n\binom{n}{d-2}\leq \frac{3\cdot 2^d}{(d-2)!}n^{d-1},
\end{equation}
where the factor $2^{d-2}$ arises since every 2-face $f$ is incident to exactly this number of cells $C$ in $\A$ and therefore $(v,f)$ can be charged by at most these many 0-borders of the form $(v,C)$.

In the complementary case, we consider 0-borders $(v,C)$ such that $C$ is a cell in $\Z$ and $v$ is a vertex of $C$ inside $K^c$, such that all the 2-faces $f$ on the boundary of $C$ that are incident to $v$, do not intersect $\partial K$ (in other words, all these 2-faces are fully contained in $K^c$). Let $(P,C)$ be a 3-border with $P$ incident to $v$, and denote by $S$ the unique 3-flat that supports $P$, which is the intersection of some $d-3$ hyperplanes of $H$. We put $H\cap S:=\{h\cap S\;|\; h\in H\;\text{and }\; S\nsubseteq h\}$, which is a collection of $n-d+3$ planes in $S$. We assign to each plane in $H\cap S$ an orientation, letting its positive side be the one that contains $P$. Then $P$ is a 3-dimensional convex polyhedron in $S$, which is the intersection of the halfspaces, defined by the positive sides of the planes in $H\cap S$. 

We want to use Euler's formula on $P$, but wish to consider only facets of $P$ that are contained in $K^c$. Let $(H\cap S)'$ be the subset of planes $h\in H\cap S$ such that $h$ supports a facet of $P$ that is contained in $K^c$. Denote by $P'$ the polyhedron in $S$, formed by the intersection of the halfspaces defined by the positive sides of the planes in $(H\cap S)'$.

By assumption, $v$ is also a vertex of $P'$, because it is incident only to facets of $P$ contained in $K^c$ and thus we did not remove any of the hyperplanes that support it. So the number of vertices of $P'$, using Euler's formula, is at most twice the number of facets of $P'$. But, by definition, this is exactly the number of facets of $P$ that are fully contained in $K^c$. (Note that the facets of $P'$ need not be fully contained in $K^c$: Each such facet $f'$ is an expansion of a facet $f$ of $P$; while $f$ is fully contained in $K^c$, $f'$ does not have to be.)

Finally, every 2-face $f$ on the boundary of $C$ is incident to exactly $\binom{d-2}{d-3}=d-2$ 3-faces $P$ on the boundary of $C$, and similarly, every vertex $v$ on the boundary of $C$ is incident to  exactly $\binom{d}{d-3}=\binom{d}{3}$ such 3-faces $P$. Therefore, the number of 0-borders $(v,C)$, with $v$ and $C$ as above, is at most
\begin{equation}\label{eq2}
\frac{2(d-2)}{\binom{d}{3}}\cdot\tau_2(K,H)=\frac{12}{d(d-1)}\tau_2(K,H).
\end{equation}

The bounds (\ref{eq1}), (\ref{eq2}), and Proposition \ref{prop:tauid} give
\begin{align*}
\tau_0(K,H)
&\leq \frac{12}{d(d-1)}\tau_2(K,H)+\frac{3\cdot 2^d}{(d-2)!}n^{d-1},\\
&\leq \left(\frac{48}{d(d-1)}+\frac{3\cdot 2^d}{(d-2)!}\right)n^{d-1}.
\end{align*}
Together with inequality (\ref{eq:tau1}), the claim of the lemma follows. $\square$
\end{prf}

Finally, returning to the proof of Theorem \ref{main}, we note that the constants of proportionality in the bounds for $\tau_0(K,H)$ and $\tau_1(K,H)$ are both at most some absolute constant (in fact, they tend to 0 as $d$ increases). This, and the bound in Proposition \ref{prop:tauid}, establish the bound in Theorem \ref{main}.~$\square$

Returning to the remark following the statement of Theorem \ref{main}, we complete the analysis to include also faces that intersect $\partial K$. We fix $0\leq k< d$, and a $k$-flat $F$ formed by the intersection of some $d-k$ hyperplanes of $H$. Every $k$-face of $\Z$ that is contained in $F$ and intersects $\partial K$ is on the boundary of exactly $2^{d-k}$ cells of $\Z$, and is a cell in the zone of $K\cap F$ in the $k$-dimensional arrangement $\A(H\cap F)$. Since the number of such cells is $\le Bn^{k-1}$, where $B$ is an absolute constant independent of $d$, and the number of $k$-flats $F$ is $\binom{n}{d-k}$, the overall number of faces of $\Z$ that intersect $\partial K$ is 
\begin{align*}
\sum_{k=0}^{d-1} Bn^{k-1}\binom{n}{d-k}2^{d-k} 
&\le Bn^{k-1}\frac{n^{d-k}}{(d-k)!}2^{d-k}\\
& = Bn^{d-1}\sum_{k=0}^{d-1}\frac{2^{d-k}}{(d-k)!}\\
& < Be^2n^{d-1}.
\end{align*}
In conclusion, we obtain the following strengthening of Theorem \ref{main}.

\begin{thm}
The overall number of faces of the cells of the zone of the boundary of an arbitrary convex set $K$ in an arrangement of $n$ hyperplane in $\R^d$, that are either fully contained in, or intersect $K^c$, is at most $Adn^{d-1}$, for some absolute constant $A>0$, independent of $d$ and of $K$. 
\end{thm}

\end{document}